\documentclass[nofootinbib,twocolumn]{revtex4}
\usepackage{epsfig,amsmath,amssymb,enumitem,graphicx,comment}
\renewcommand{\vec}[1]{\boldsymbol{\mathrm{#1}}}
\newcommand{\be}{\begin{equation}}
\newcommand{\ee}{\end{equation}}

\begin{document}
\title{Applying Modified Gravity (MOG) to the Lensing and Einstein Ring in Abell 3827}
\author{J. W. Moffat${}^{1,2}$ and V. T. Toth${}^1$\\~\\
${}^1$Perimeter Institute for Theoretical Physics, Waterloo, Ontario N2L 2Y5, Canada\\
and\\
${}^2$Department of Physics and Astronomy, University of Waterloo, Waterloo,\\
Ontario N2L 3G1, Canada}




\begin{abstract}
The lensing and Einstein ring at the core of the galaxy cluster Abell 3827 are reproduced in the modified gravity theory MOG.
The estimated effective lensing mass $M_L=(1+\alpha)M_b=5.2\times 10^{12} M_\odot$ within $R=18.3$~kpc for a baryon mass $M_b=1.0\times 10^{12} M_\odot$ within the same radius produces the observed Einstein ring angular radius $\theta_E=10''$. A detailed derivation of the total lensing mass is based on modeling of the cluster configuration of galaxies, intracluster light and X-ray emission. The MOG can fit the lensing and Einstein ring in Abell 3827 without dark matter as well as General Relativity with dark matter.
\end{abstract}
\maketitle

\section{Introduction}

Dark matter has not been observed in laboratory experiments~\cite{Baudis}. Therefore, it is important to consider a modified gravitational theory. The difference between standard dark matter models and modified gravity is that dark matter models assume that General Relativity (GR) is the correct theory of gravity and a dark matter particle such as WIMPS, axions and fuzzy dark matter are postulated to belong to the standard particle model. The present work using modified gravity (MOG) reproduces the basic lensing and Einstein ring around the center of the galaxy cluster Abell 3827~\cite{Chen2020}. The Einstein ring is made of multiple lensed images of a background spiral galaxy modelled by a lensing mass distribution. In Abell 3827 there are four bright and dominant cluster member galaxies along with a dimmer member galaxy within the ring, an intracluster light and an extended X-ray emission centered on the ring. We analyze the lensing and Einstein ring with a modified gravity MOG, also called Scalar--Tensor--Vector--Gravity (STVG)~\cite{Moffat2006,Moffat2020}. The MOG is described by a fully covariant action and field equations, extending GR by the addition of two gravitational degrees of freedom. The first is $G= 1/\chi$, where $G$ is the coupling strength of gravity and $\chi$ is a scalar field. The second degree of freedom is a massive gravitational vector field $\phi_\mu$. The gravitational coupling of the vector graviton to matter is universal with the gravitational charge $Q_g=\sqrt{\alpha G_N}M$, where $\alpha$ is a dimensionless function of the scalar field $\chi$, $G_N$ is Newton's gravitational constant and $M$ is the mass of a body.

We write the gravitational coupling coefficient $G$ as $G=G_N(1+\alpha)$. The effective running mass of the spin-1 vector graviton is determined by the parameter $\mu$, which fits galaxy rotation curves and cluster dynamics without exotic dark matter~\cite{BrownsteinMoffat2007,MoffatRahvar2013,MoffatRahvar2014,IsraelMoffat2018,DavariRahvar2020,MoffatHaghighi2017}. It has the value $\mu\sim 0.01 - 0.04\,{\rm kpc}^{-1}$, corresponding to $\mu^{-1}\sim 25 - 100$ kpc and an effective mass $m_\phi\sim 10^{-28}\,{\rm eV}$. MOG has been used to model cosmology~\cite{MoffatToth2013,JamilaRoshan2016,Moffat2020,Landau2020}. A detailed derivation of the lensing properties and the Einstein ring of a massive object has been published~\cite{MoffatToth2009a,RahvarMoffat2019,MoffatRahvarToth2018}.

The total {\em effective} lensing mass $M_L$ is approximated by a central source mass within the Einstein ring with the angular radius $R_E=10''$ (18.3 kpc at z=0.099), and the fitting of MOG to the Einstein ring gives the lensing mass estimate $M_L\sim 5.2\times 10^{12} M_\odot$ consistent with the estimate from models of the central region of the galaxy cluster $M_L\sim 4.5\times 10^{12} M_\odot$~\cite{Chen2020}, with a baryon mass $M_b=1.0\times 10^{12} M_\odot$.

\section{The MOG Field Equations}

We introduce $\chi= 1/G$ where $\chi$ is a scalar field and $G=G_N(1+\alpha)$ is the coupling strength of gravity, expressed in terms of $\alpha=(\chi G_N)^{-1}-1$.
The field equations are given by (we use the metric signature $(+,-,-,-)$ and units with $c=1$)~\cite{Moffat2006,Moffat2020}:
\begin{align}
\label{Gequation}
G_{\mu\nu}={}&-\frac{\omega_M}{\chi^2}\biggl(\nabla_\mu\chi\nabla_\nu\chi -\frac{1}{2}g_{\mu\nu}\nabla^\alpha\chi\nabla_\alpha\chi\biggr)\nonumber\\
&-\frac{1}{\chi}(\nabla_\mu\chi\nabla_\nu\chi-g_{\mu\nu}\Box\chi)+\frac{8\pi}{\chi}T_{\mu\nu},
\end{align}
\be
\label{Bequation}
\nabla_\nu B^{\mu\nu}+\mu^2\phi^\mu=J^\mu,
\ee
\be
\label{Boxchi}
\Box\chi=\frac{8\pi}{(2\omega_M+3)}T,
\ee
where $G_{\mu\nu}=R_{\mu\nu}-\frac{1}{2}R$, $\nabla_\mu$ denotes the covariant derivative with respect to the metric $g_{\mu\nu}$, $B_{\mu\nu}=\partial_\mu\phi_\nu-\partial_\nu\phi_\mu$ and $\omega_M$ is a constant. Moreover, $\Box=\nabla^\mu\nabla_\mu$, $J^\mu=\sqrt{\alpha G_N}\rho u^\mu$, $\rho$ is the density of matter and field energy and $u^\mu=dx^\mu/ds$. $\Lambda$ is the cosmological constant and $\mu$ is the effective running mass of the spin 1 graviton vector field. The energy-momentum tensor is
\be
T_{\mu\nu}=T^M_{\mu\nu}+T^\phi_{\mu\nu}+g_{\mu\nu}\frac{\chi\Lambda}{8\pi},
\ee
where
\be
T^\phi_{\mu\nu}=-\biggl({B_\mu}^\alpha B_{\alpha\nu}-\frac{1}{4}g_{\mu\nu}B^{\alpha\beta}B_{\alpha\beta}+\mu^2\phi_\mu\phi_\nu-\frac{1}{2}\mu^2g_{\mu\nu}\phi^\alpha\phi_\alpha\biggr),
\ee
and $T=g^{\mu\nu}T_{\mu\nu}$.

The equation of motion for a massive test particle in MOG has the
covariant form~\cite{Moffat2006,Roshan2013}:
\begin{equation}
\label{eqMotion}
m\biggl(\frac{du^\mu}{ds}+{\Gamma^\mu}_{\alpha\beta}u^\alpha
u^\beta\biggr)= q_g{B^\mu}_\nu u^\nu,
\end{equation}
where ${\Gamma^\mu}_{\alpha\beta}$ denote the Christoffel
symbols. Moreover, $m$ and $q_g$ denote the test particle mass $m$ and
gravitational charge $q_g=\sqrt{\alpha G_N}m$, respectively.  For a
massless photon the gravitational charge vanishes,
$q_\gamma=\sqrt{\alpha G_N}m_\gamma=0$, so photons travel on null
geodesics $k^\nu\nabla_\nu k^\mu=0$~\cite{GreenMoffatToth2018}:
\begin{equation}
\frac{dk^\mu}{ds}+{\Gamma^\mu}_{\alpha\beta}k^\alpha k^\beta=0,
\end{equation}
where $k^\mu$ is the photon momentum and $k^2=k^\mu k_\mu=0$. We
note that for $q_g/m=\sqrt{\alpha G_N}$ the equation of motion for a
massive test particle (\ref{eqMotion}) {\it satisfies the (weak)
equivalence principle}, leading to the free fall of particles in a
homogeneous gravitational field, although the free-falling particles
do not follow geodesics.

A phenomenological formula for $\alpha$ for approximately constant $\alpha$ and weak gravitational field is~\cite{MoffatToth2009b}:
\be
\label{alphaformula}
\alpha=\alpha_{\infty}\frac{M}{(\sqrt{M}+E)^2},
\ee
where $\alpha_{\infty}\sim{\cal O}(10)$ and 
$E\lesssim{\cal O}(10^5)$.
For $r\ll \mu^{-1}\sim 25 -100$ kpc the MOG acceleration reduces to the Newtonian acceleration. 

The modified Newtonian acceleration law for a point particle can be written
as~\cite{Moffat2006}:
\begin{equation}
\label{MOGaccelerationlaw}
a_{\rm MOG}(r)=-\frac{G_NM}{r^2}[1+\alpha-\alpha\exp(-\mu r)(1+\mu r)].
\end{equation}
This reduces to Newton's gravitational acceleration in the limit
$\mu r\ll 1$. This is consistent with
solar system observational data.

In the limit that $r\rightarrow\infty$, we get from
(\ref{MOGaccelerationlaw}) for approximately constant $\alpha$ and
$\mu$:
\begin{equation}
\label{AsymptoticMOG}
a_{\rm MOG}(r)\approx -\frac{G_N(1+\alpha)M}{r^2}.
\end{equation}
The MOG acceleration has a Newtonian--Kepler behavior for large $r$ with enhanced
gravitational strength $G=G_N(1+\alpha)$. The transition from
Newtonian acceleration behavior for small $r$ to non-Newtonian
behavior for intermediate values of $r$ is due to the repulsive
Yukawa contribution in (\ref{MOGaccelerationlaw}). This can also
result in the circular orbital rotation velocity $v_c$ having a
maximum value in the transition region.

We have written the gravitational strength $G=G_N(1+\alpha)$, so we can relate the scalar field $\chi$ to this expansion of $G$ with $\alpha$ as $\chi\sim 1/G_N(1+\alpha)$. The constant $\omega_M$ can be set to $\omega_M=1$. In the weak gravitational acceleration formula (\ref{MOGaccelerationlaw}), the acceleration of a particle identified with a planet in the solar system approaches the Newtonian acceleration law as $r$ approached the size of the solar system.
In Brans-Dicke theory~\cite{BransDicke}, the constant $\omega$ is chosen to be large so that the theory can agree with solar system measurements. An important difference between Brans-Dicke gravity and MOG is the additional degree of freedom of the gravitational vector field $\phi_\mu$, allowing us to obtain the weak field MOG acceleration formula (\ref{MOGaccelerationlaw}). We have adopted a different approach for obtaining the solar system weak field limit by choosing the equivalent constant $\omega_M=1$ and the relation $\chi=1/G_N(1+\alpha)$. In the fitting of data when applying the field acceleration formula (\ref{MOGaccelerationlaw}) to {\it weak gravitational fields}, the parameters $\alpha$ and $\mu$ are treated as running constants that are not universal constants. The magnitude of $\alpha$ depends on the physical length scale or averaging scale $\ell$ of the system. For the solar system, $\ell_\odot\sim 0.5$ pc and for a galaxy $\ell_G\sim 5-24$ kpc.

\section{MOG Lensing and the Einstein Ring}

The MOG formula for light bending for weak gravitational fields is given by the deflection~\cite{Moffat2006,MoffatRahvarToth2018}:
\be
\theta_L=\frac{4GM}{r_0}=\frac{4G_N(1+\alpha)M}{r_0}.
\ee
The bending of light predicted by MOG is equal to the bending of light predicted by GR using the effective lensing mass:
\be
\label{Lensing mass}
M_L=(1+\alpha)M.
\ee

The data for the projected total lensing mass is based on the projected mass from the four brightest cluster galaxies and the intracluster gas~\cite{Chen2020}. The standard lensing equation is given by
\be
r_{BS}=r_S-\theta_L(r_S,\Sigma(r_S)),
\ee
where $r_{BS}$ is the observed position of the background source, $r_S$ is the observed position of the source, and
$\Sigma(r_S)$ is the projected surface mass density of the cluster at the position $r_S$. In the thin lense approximation, the projected mass density is given by
\be
\Sigma({\vec r})=\int dz\rho({\vec r},z),
\ee
and the deflection angle is
\be
\theta_L({\vec r})=\frac{4G}{c^2}\int d^2r'\frac{\Sigma({\vec r})({\vec r}-{\vec r}')}{|{\vec r}-{\vec r}'|^2}.
\ee

When light from a distant source is deflected by a massive object, an observer aligned with the distant source and the massive object may see a ring-like image of the distant source called an Einstein ring. If the distances to the remote object and the massive object are known, the apparent size of the Einstein ring can be used to estimate the effective lensing mass. If dark matter is present, then the effective lensing mass is $M_L=M_b+M_{DM}$, whereas if dark matter is absent, the effective lensing mass is determined by the baryon mass $M_b$. The apparent radius $\theta_E$ of an Einstein ring is in radians~\cite{Weinberg,MoffatRahvarToth2018}:
\be
\theta_E=\biggl(\frac{4G_NM_L}{c^2}\frac{d_{LS}}{d_Ld_S}\biggr)^{1/2},
\ee
where $d_L,.d_S$ and $d_{LS}$ are the angular diameter distances, related to the comoving distance $D$ by $d=D/(1+z)$ for a flat, $\Omega=1$, cosmology, to a lensing object, the source, and between the lensing object and the source, respectively.

For an effective lensing mass $M_L=5.2\times 10^{12} M_\odot$, $d_L=377$ Mpc, $d_S$=1718 Mpc, and $d_{LS}$=1533 Mpc, we get the Einstein ring angular radius:
\be
\theta_E=10''.
\ee
The quoted observational angular Einstein ring radius is $\theta_E=10''$, corresponding to 18.3 kpc. Models of the projected total lensing mass with dark matter inside the Einstein radius give $M_L\sim 4.5\times 10^{12} M_\odot$~\cite{Chen2020}.

\begin{figure}
\includegraphics[width=\linewidth]{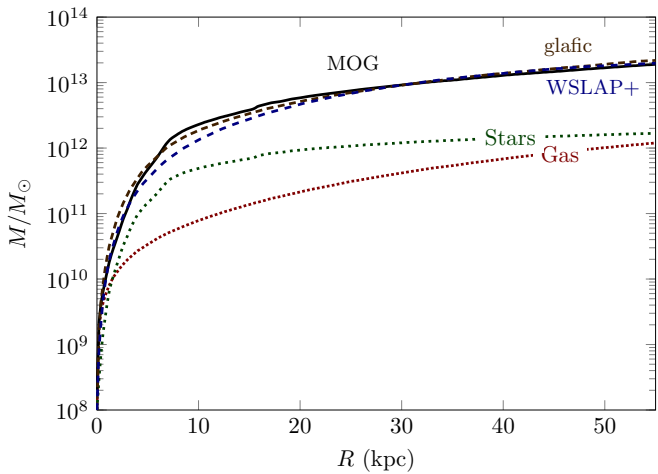}
\caption{Projected cumulative mass within radius $R$ from the cluster center, adapted from \cite{Chen2020}. Stellar (green, dotted) and gas (red, densely dotted) curves are shown, in addition to the ``WSLAP+'' (blue, dashed) and ``glafic'' (brown, densely dashed) models discussed in that reference. The solid black curve shows the MOG prediction using the parameters described in the text.\label{fig:mass}}
\end{figure}

We can develop a more detailed analysis of the lensing and Einstein ring by invoking the spatial dependence of the scalar gravitational coupling strength $G_N(1+\alpha)=1/\chi$. The density profile can be expressed as
\be
\label{density}
\rho(r)=(1+\alpha(r))\rho_b,
\ee
where
\be
\rho_b=\rho_*+\rho_g,
\ee
and $\rho_*$ and $\rho_g$ denote the stellar and gas densities, respectively. We can model the total lensing mass $M_L(r)$ as the three-dimensional average:
\be
M(r)=4\pi\int_0^rdr'r^{'2}\rho(r)=4\pi\int^r_0 dr'r^{'2}(1+\alpha(r'))\rho_b(r').
\ee
We model the radial dependence of $\alpha(r)$ using Eq.~(\ref{alphaformula}) as follows:
\be
\alpha(r)=\alpha_{\infty}\frac{M(r)}{(\sqrt{M(r)}+E)^2}.
\ee
A good fit to the WSLAP+ and glafic models is displayed in Fig.~\ref{fig:mass}. This is achieved using the parameters $\alpha_{\infty}=11$, $E=680000~M_\odot^{1/2}$, but we note that the dependence is near degenerate: fits almost as good are obtained using, e.g., $\alpha_{\infty}=4$, $E=25000~M_\odot^{1/2}$ or $\alpha_{\infty}=19$, $E=1.2\times 10^6~M_\odot^{1/2}$.

Within 18.3~kpc, our best fit model of $\alpha(r)$ yields $\alpha(18.3~{\rm kpc})\sim 4.0$, yielding an effective lensing mass of
\be
M_L=(1+\alpha)M_b=5.2\times 10^{12}~M_\odot,
\ee
using the sum of the enclosed stellar and gas masses, $M_b\sim 1.0\times 10^{12} M_\odot$ as the baryonic mass.

The value $\alpha_{\infty} \sim 11$ is consistent with the $\alpha$ determined by fitting MOG to clusters~\cite{BrownsteinMoffat2007,IsraelMoffat2018,MoffatRahvar2014,MoffatHaghighi2017}. It is also consistent with the MOG fitting of spiral galaxies~\cite{MoffatRahvar2013}.

\section{Conclusions}

The modified gravity theory MOG can reproduce the lensing and the Einstein ring for the galaxy cluster Abell 3827, using a central mass approximation in estimating the lensing mass inside the Einstein ring. The MOG lensing mass $M_L=(1+\alpha)M_b$ is $M_L\sim 5.2\times 10^{12} M_\odot$ and the estimated baryon mass obtained in Ref.~\cite{Chen2020} is $M_b\sim 1.0\times 10^{12} M_\odot$. This gives the Einstein ring angular radius $R_E=10''$ with $\alpha_\infty={\cal O}(10)$, which is consistent with the fitting of rotation velocities of spiral galaxies~\cite{MoffatRahvar2013} and clusters~\cite{BrownsteinMoffat2007,IsraelMoffat2018,MoffatRahvar2014,MoffatHaghighi2017}. The Einstein ring in Abell 3827 can be fitted with a dark matter halo model. The dark matter component has the same orientation as the light from the intracluster stars, suggesting that the intracluster stars trace the gravitational potential of this component~\cite{Chen2020}. We have fitted the galaxy, ICL and X-ray emission configuration of the cluster to the data by modeling the spatial dependence of the scalar field $\alpha(r)$. In the absence of dark matter, there is agreement between the projected sky distribution of the lensing mass and the visible baryon mass.
The consistency of the radial mass profile with the WSLAP+ and glafic models that, in turn, were used to successfully model the lensing of Abell 3827 \cite{Chen2020} suggests that MOG, too, can be used to develop a good two-dimensional fit. This analysis is yet to be performed and, when completed, will be published elsewhere.

\section*{Acknowledgments}

We thank Mandy Chen and Martin Green for helpful discussions. Research at the Perimeter Institute for Theoretical Physics is supported by the Government of Canada through industry Canada and by the Province of Ontario through the Ministry of Research and Innovation (MRI).

\end{document}